\documentclass[aps,prx,reprint,longbibliography,superscriptaddress,floatfix]{revtex4-2}
\usepackage{graphicx}
\usepackage{xcolor}
\usepackage{amsmath}
\usepackage{url}
\begin{document}

\title{Unmodeled states and uncertain action outcomes in agentic scanning tunneling microscopy}
\author{Siyu Cheng}
\email[Contact author: ]{siyu.cheng@bc.edu}
\author{Muxian Xu}
\author{Christopher Candelora}
\author{Ilija Zeljkovic}
\email[Contact author: ]{ilija.zeljkovic@bc.edu}
\affiliation{Department of Physics, Boston College, Chestnut Hill, Massachusetts 02467, USA}
\date{October 22, 2026}

\begin{abstract}
In physical experiments, interventions can alter hidden experimental states in ways that cannot be predicted in advance. Autonomous scientific agents must therefore interpret the consequences of their interventions while operating with incomplete knowledge of the experimental state.
Here we investigate this problem using scanning tunneling microscopy (STM) tip conditioning, traditionally a human-expert-intensive task governed by inaccessible tip apex conditions and uncertain action outcomes. We introduce \texttt{quailbot}, an agent harness that places a large language model (LLM) inside the instrument feedback loop by linking physical interventions with experimental readbacks. After a brief apprenticeship with a human expert, frontier LLM agents autonomously completed end-to-end tip conditioning on a low-temperature STM using tungsten tips on FeTe at 4.5~K and passed an independent verification. The contingencies that arose during the experiments further exposed the limits of LLM agentic autonomy when relevant experimental states or dynamics were hidden or unmodeled. Our results show that future autonomous physical experimentation requires agents to infer hidden and unmodeled experimental states, track the outcomes of their actions, and operate within the observability and action limits of real instruments.
\end{abstract}

\maketitle

\section{Introduction}

Automation is reshaping experimental science, from robotic materials synthesis~\cite{szymanski2023alab} to closed-loop optimization of complex devices such as spin qubits~\cite{schuff2026spinqubit}. Large language model (LLM) agents mark a new stage in experimental automation. The interface between the agent and the instrument is growing and becoming a public standard~\cite{anthropic2026mhs}. Given tool access to real instruments, the agents plan and execute chemical syntheses~\cite{boiko2023coscientist,bran2024chemcrow}, run multistage accelerator measurements~\cite{hellert2026accelerator}, characterize superconducting qubits~\cite{li2026qubit}, operate electron microscopes and control experiments in materials laboratories~\cite{wall2026temagent,shi2026qumus}. Much of this has been demonstrated in scenarios in which the relevant experimental states and operations are sufficiently well specified for the agent to plan around them. Recent work has begun to tackle the challenges of departures from these nominal experimental conditions. An agent has navigated a cryogenic microwave impedance microscope under uncertain perception, sample inhomogeneity, and ambiguous physics~\cite{qiu2026aims}, while an X-ray agent identified and corrected an unexpected motor offset during autonomous sample alignment~\cite{xray2026scientist}. However, a distinct source of uncertainty, which we define as action-outcome uncertainty, remains underexplored. The agent's  intervention can change the physical system in an unpredictable manner, which can only become apparent through subsequent measurements. This challenge brings a crucial question -- what should an agent do when an action intended to improve the system may instead degrade it?

\begin{figure*}[!t]
\centering
\includegraphics[width=1.0\textwidth]{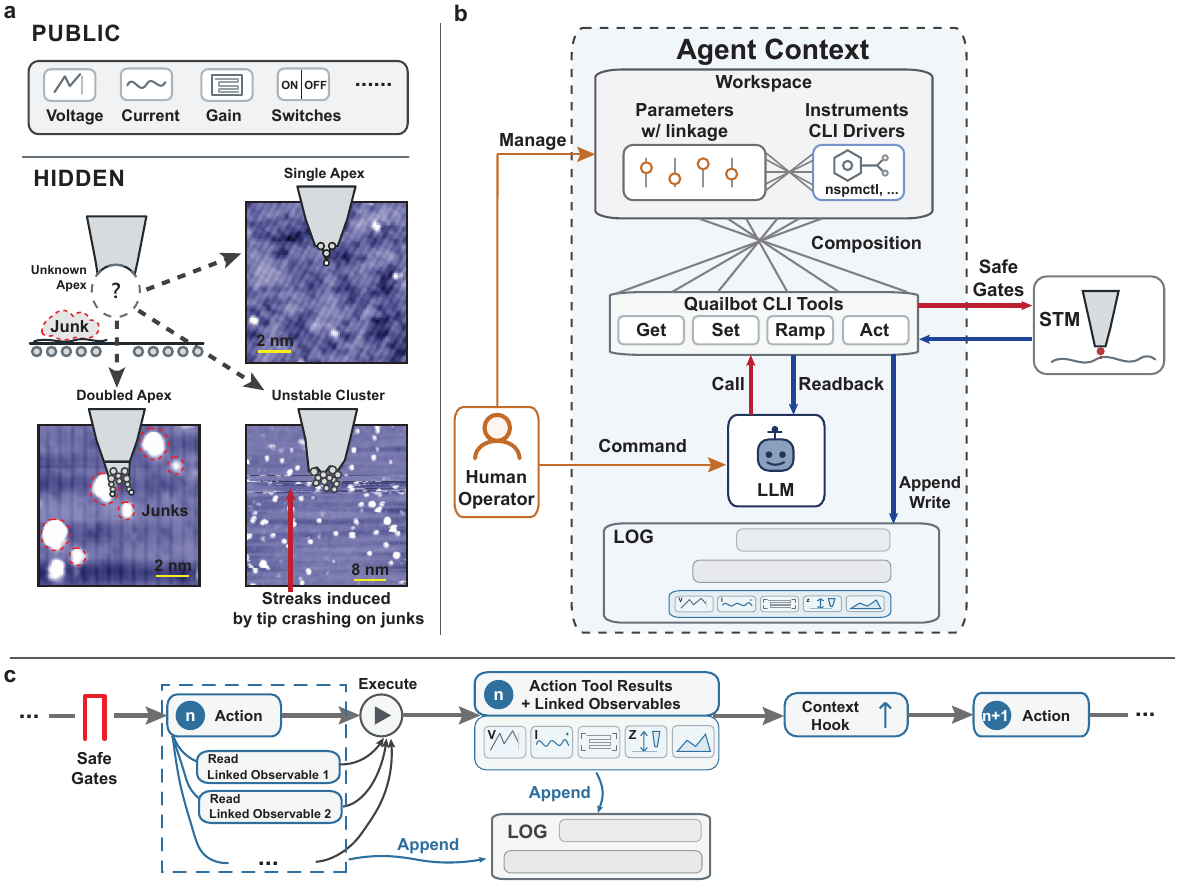}
\caption{\textbf{Overview of hidden states, tip conditioning and the \texttt{quailbot} harness.}
 (a), A representative example of public states and hidden states in tip conditioning. Tip conditioning acts on an unknown apex and can produce distinct imaging outcomes. The three topographs show a single apex with lattice contrast, a doubled apex with junk, and an unstable cluster with streaks.
 (b), The overview of the \texttt{quailbot} harness. The human operator defines the workspace and instrument interfaces, including the linked observables of each declared capability. The LLM selects declared capabilities and supplies their arguments through four instrument-control CLI tools: Get, Set, Ramp and Act. The harness returns tool results to the LLM, including linked readbacks after state-changing calls and appends each record to the experiment log.
 (c), Breakdown of one state-changing call. After the call passes the safe gates, the harness executes it and reads the linked observables defined for that capability in the workspace. The action result and linked-observable results are appended to the experiment log as mentioned in (b). The context hook injects these results into the agent context before the agent selects the next call. STM setup conditions: (a) (Single Apex and Unstable Cluster), $V_{sample}$ = 100 mV, $I_{set}$ = 200 pA; (a) (Doubled Apex), $V_{sample}$ = 100 mV, $I_{set}$ = 300 pA.}
\label{fig:harness}
\end{figure*}

This action-outcome uncertainty can be described in terms of the information available at the agent-instrument interface, and we use the term "state" operationally from this perspective. The public states are information directly exposed through the instrument interface, such as settings and readbacks. Hidden states are experimentally relevant physical information that influence observations but are not directly exposed. Some hidden states are included in the agent's working model even though they are not directly measurable, and may still be inferred from suitable measurements. Others are unfamiliar to the model -- unmodeled latent contingencies that may enter the working representation only when observations become inconsistent with the assumed state of the experiment. We use "hidden dynamics" to describe the incompletely understood processes through which interventions change these states. Partial observability and uncertain state transitions are established problems in sequential decision making. Our focus is their experimental manifestation when an LLM directs a real scientific instrument. A scanning tunneling microscope (STM) provides a direct physical realization of this anatomy (Fig.~\ref{fig:harness}(a)). Bias voltage, current setpoint, and scan frame are public states. The atomic configuration of the tip apex is the hidden state that is experimentally decisive but cannot be observed directly and must instead be inferred from the measurements it produces. The tip-conditioning interventions act on this hidden state. A voltage pulse or controlled tip-sample contact can sharpen the apex, dull it, or pick up an atomic cluster, without a deterministic way to know beforehand which outcome will occur.

Tip conditioning prepares the tip to produce reliable STM images and spectroscopic data, and in many STM experiments it must be completed before any meaningful measurements can begin. Some STM measurements further require a spin-polarized tip, which enables spin-resolved atomic-scale imaging~\cite{Wiesendanger2009SpinScale,enayat2014spstm,Zhao2019,sharma2023fete}, and adds another hidden property to the tip state. Tip conditioning can change the apex geometry, the magnetic sensitivity of the tip, and the stability of the tunneling junction~\cite{singh2015magnetictips,trainer2019feTe}. However, the tip conditioning process is laborious, relies heavily on trial and error, and can consume hours of expert judgment. The operator must simultaneously explore the sample and reshape the probe, infer from indirect evidence whether a streaky STM image originates from surface debris or an unstable tip, or whether paired features are real surface structures or a double-apex artifact. Automation of this process has therefore long been sought, although many autonomous-exploration frameworks still leave conditioning as a separate procedure~\cite{kalinin2021autonomous,narasimha2024bo,narasimha2025activelearning}. Recent automation systems infer tip quality with learned neural network classifiers and close conditioning loops using trained task-specific control policies~\cite{gordon2019tipstate,diao2024smallsystems,rashidi2018tip,krull2020deepspm,zhu2024stmimaging,wang2021tipconditioning}. A remaining challenge is how automation should respond when an experiment does not proceed as expected. Language-model agents offer a distinct route. Instead of relying on a pre-engineered policy specialized to a predefined state and action space, LLM agents can interpret new measurements using prior experimental knowledge and the history of their actions. They can also incorporate information supplied during the experiment and revise their strategy accordingly. LLMs for scanning-probe automation have already progressed from natural-language command translation to high-level planning and orchestration of task-specific SPM automation modules~\cite{diao2024llmspm,liu2024microscopy,mandal2025aila,diao2026integrating}.

Here we report end-to-end LLM tip conditioning autonomy on a low-temperature STM, using tungsten tips on FeTe at 4.5~K. The LLM agents operated through \texttt{Quailbot} ("quail" stands for Quantum Uncertain Action-Outcome Instrument Loop), an agent harness~\cite{young2025harnesses} we developed that allows LLMs to operate real instruments. The agent developed procedural knowledge through human-guided apprenticeship. Two launch-prompt-only runs delivered measurement-ready tip endpoints without further human intervention. We also examined how they responded to unexpected experimental conditions. These cases showed both autonomous adaptation and failures to identify the underlying problem, and revealed where additional human information or physical intervention was needed. These findings motivate a closer examination of the diagnostic information and physical capabilities available to agents through the instrument interface.

\section{Results}
\subsection{Agent harness and apprenticeship}
The agents operated the STM through \texttt{Quailbot}. The human operator manages a workspace that defines the available instrument capabilities and the measurements linked to them (Fig.~\ref{fig:harness}(b)). The LLM accesses these capabilities through four instrument-control CLI tools. Get reads a parameter, Set writes a value, Ramp steps a parameter between values and Act invokes a named action. A state-changing call proceeds as follows (Fig.~\ref{fig:harness}(c)). The agent selects Set, Ramp or Act and supplies a declared capability and its arguments. The harness checks the call against the workspace before the CLI command reaches the instrument. After execution, the harness retrieves the linked observables defined in the workspace for that capability. The harness returns the tool result and the linked measurements to the agent before it decides its next action. The harness also appends this record to the experiment log. The complete execution and logging procedure is described in Methods. In this work we used gpt-5.6-sol and claude-opus-5 in the \texttt{quailbot} harness to run the tip conditioning task on our STM, with gpt-5.4-mini as a deliberately lower intelligent probe.

The harness also includes memory and skills modules to enable the LLM's continuous learning~\cite{sumers2024CoALA}. The human operator trained the agent as an apprentice on how to run the STM, and the agent consolidated this training using its memory and skills modules. Initially, the agent had to request human permission before taking actions, but it was later allowed to operate more autonomously. Human instruction covered conditioning procedures such as tip shaking and bias pulsing. Tip shaking is a repetitive controlled tip punching on the sample surface that is used to reshape the apex. Bias pulsing applies a pulsed voltage across the junction. The agent was then taught domain-specific judgment, such as how to distinguish surface contamination from tip-related imaging artifacts. The operator monitored the instrument GUI and agent transcripts and provided corrections during the apprenticeship.

The instruction also covered experimental consequences that were not obvious from the action readbacks alone. During the subsequent graduation run, for example, the operator corrected the agent’s use of a ramp to reach the shaking amplitude, because the ramp made each shake far longer than what agent planned. The graduation run reached a single, stable, atom-resolving tip endpoint but still required two human steering messages. The teaching record, consolidation history and run-specific knowledge versions are documented in the Supplemental Material~\cite{SM}. Memory and skill mechanisms are described in Methods.

\begin{figure*}[!t]
\centering
\includegraphics[width=0.92\textwidth]{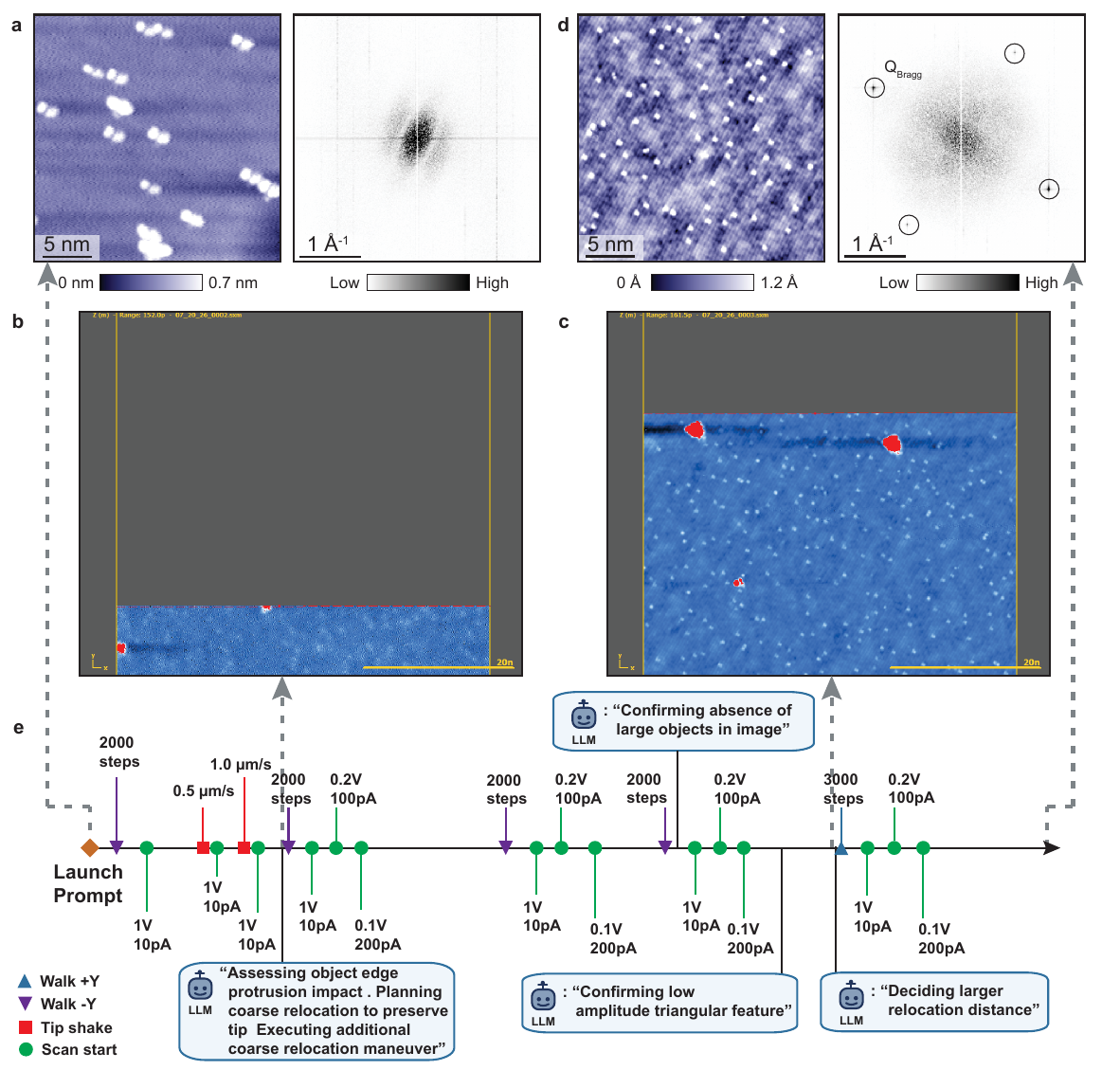}
\caption{\textbf{Anatomy of an end-to-end STM tip-conditioning run.}
(a), STM topograph on FeTe acquired before the launch and its corresponding FFT. (b), Partial topograph from the linked live-scan observable, showing a protrusion at the object edge. The scan was stopped by the agent before the topograph completed. (c), Partial topograph from the linked live-scan observable, showing a triangular feature and a local streak. (d), The final delivered topograph and its corresponding FFT. The four circles mark the first-order Bragg peaks. The BraggZ scores are 20.7 in (a) and 295.4 in (d). (e), Chronological trajectory of the presented run. Task events, agent-issued commands, and representative agent reasoning summaries are marked on the timeline. Events for panels (a)--(d) are indicated by the dashed arrows. STM setup conditions: (a), $V_{sample}$ = 1000 mV, $I_{set}$ = 10 pA; (d), $V_{sample}$ = 100 mV, $I_{set}$ = 200 pA.}
\label{fig:launchonly}
\end{figure*}

\begin{figure*}[!t]
\centering
\includegraphics[width=0.88\textwidth]{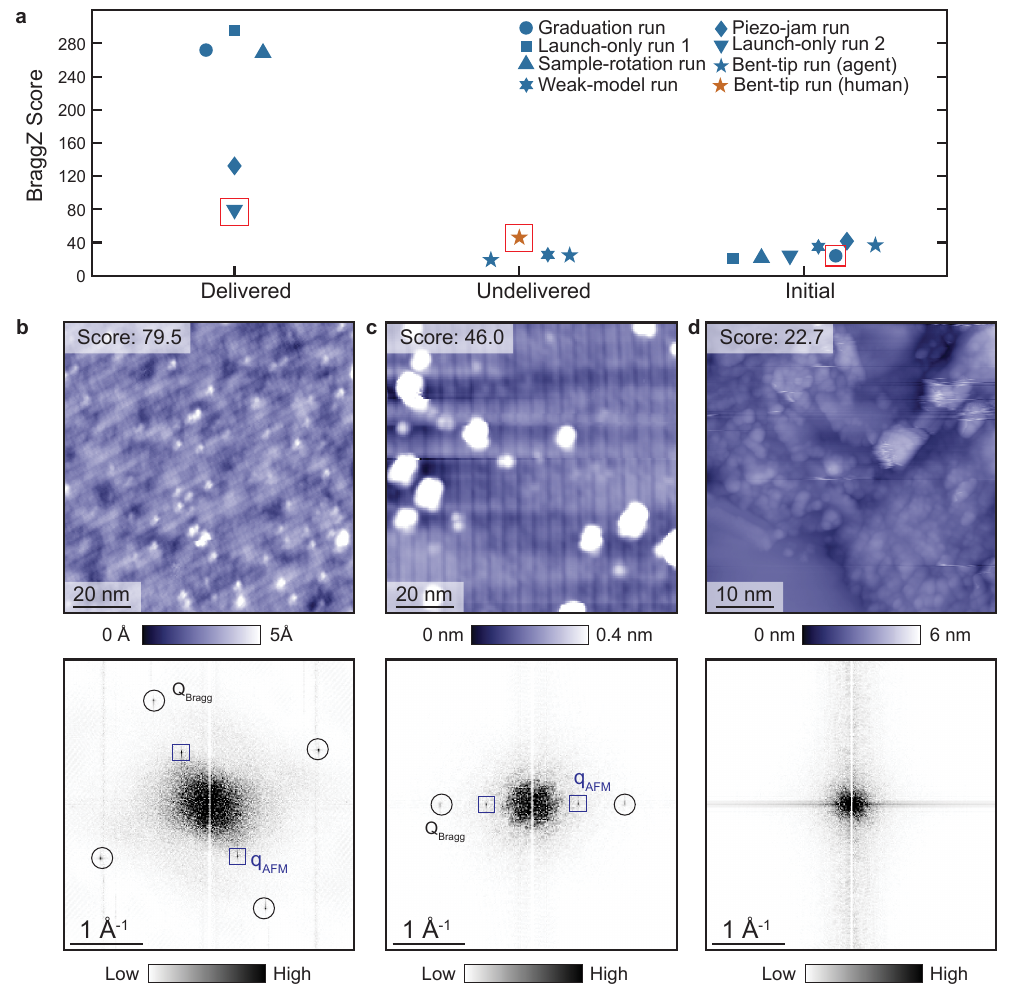}
\caption{\textbf{Post-run assessment of initial and endpoint STM images.}
(a), BraggZ scores of topographs across all runs discussed in this paper. Topographs are classified as delivered, undelivered and initial states. The "delivered/undelivered" classification is based on human expert's assessment. The "initial" states are topographs taken before a run starts. Blue markers denote agent-run results and the orange marker denotes the result after human takeover. Red boxes highlight the examples in (b)--(d).
(b), Delivered topograph and corresponding FFT from Launch-only run 2, with a BraggZ score of 79.5. Four first-order Bragg peaks are resolved. The tip remains stable across isolated small surface protrusions that produce enhanced low-wavevector intensity in the FFT.
(c), Undelivered topograph and corresponding FFT after the human operator's best-effort conditioning of the intentionally bent tip, with a BraggZ score of 46.0. The tip crashed into large surface protrusions and produced noticeable scan streaks. Bragg peaks are resolved along only one direction. In (b) and (c), black circles mark the first-order Bragg peaks $Q_{\mathrm{Bragg}}$. Blue boxes mark the $q_{\mathrm{AFM}}$ peaks of the $2a_0$ spin texture imaged with a spin-polarized tip in (b) and (c), where $a_0$ is the lattice constant.
(d), Initial topograph and corresponding FFT from the graduation run. The topograph shows strong corrugation and numerous tip-crash streaks.
STM setup conditions: (b), $V_{sample}$ = 100 mV, $I_{set}$ = 200 pA; (c),(d), $V_{sample}$ = 100 mV, $I_{set}$ = 300 pA.}
\label{fig:verification}
\end{figure*}

\begin{figure*}[!t]
\centering
\includegraphics{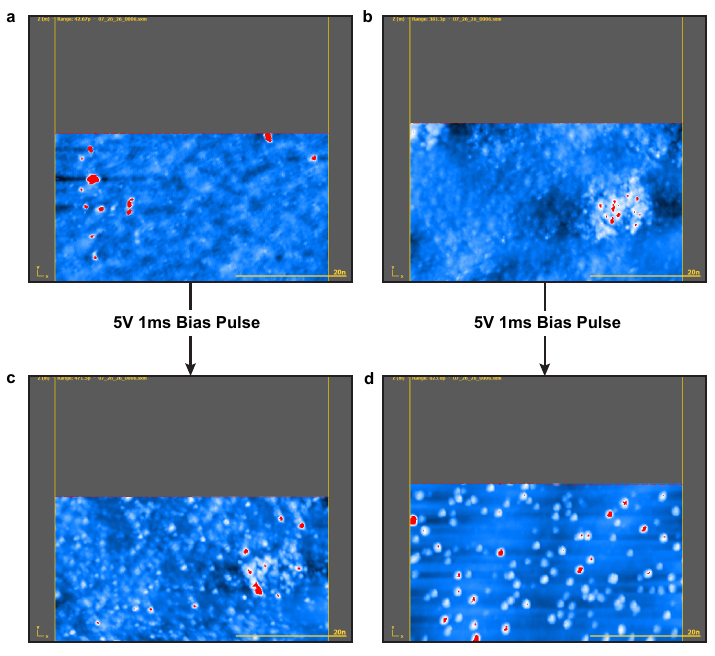}
\caption{\textbf{Bias pulse interventions in launch-only run.}
(a)--(d), Live-scan images from two successive bias pulse interventions of the same $50\times50$~nm field of view in the claude-opus-5 launch-only run. (a) and (b), Images before each bias pulse intervention. (c) and (d), Images after each corresponding bias pulse intervention. For each intervention, the pulse was performed at a corner site that was 216~nm away from the imaging field of view, and the tip was returned to the fixed field of view for the before and after scans. Both pulses used 5~V and 1~ms.}
\label{fig:interventions}
\end{figure*}

\subsection{Launch-only delivery and post-run assessment}
In the launch-only runs, the agent demonstrated the ability to deliver the task end to end. The agent received a task prompt and then operated without further human messages until its terminal report. The STM tip was deliberately sabotaged before the task. The first launch-only run finished in 70.8 minutes. The pre-launch topograph showed the tip had doubled apex and no lattice resolved (Fig.~\ref{fig:launchonly}(a)). The agent then conditioned the tip with a gentler variant of the taught gain shake, but with its own decision on the parameter choice, capping the integral gain at 1.0~\textmu m/s (Fig.~\ref{fig:launchonly}(e)). To compare, during the apprenticeship, the human operator only taught the agent to use 10~\textmu m/s.
The agent repeatedly made new scans to observe what it did. These readbacks changed the agent's plan. When a scan showed a protrusion rising toward the tip (Fig.~\ref{fig:launchonly}(b)), the agent withdrew the tip instead of continuing. Later, in a field of view where it expected to see a clean area, the completed scan showed a triangular junk feature. This led the agent to change its strategy again (Fig.~\ref{fig:launchonly}(c)). The agent increased the piezo walk step by 50\% (3000 motor steps along Y). Finally, it reached a new site and reported the task finished with a 50~nm, 448-pixel topograph at 100~mV and 200~pA with the atomic lattice resolved (Fig.~\ref{fig:launchonly}(d)).

The agent read the live scanned topograph and assessed whether the tip was ready for measurement. A measurement-ready tip usually produces a clear-lattice-resolving topograph. We verified the delivered tip endpoints post hoc using the raw topographs, their Fourier transforms and the BraggZ metric described in Methods. A human STM expert performed the verification, rather than relying on the agent's completion report. BraggZ was used as a quantitative measure of lattice-peak contrast. Fig.~\ref{fig:verification}(a) places the BraggZ score of every endpoint in runs discussed in this paper. The BraggZ scores of agent delivered states spanned from 79.5 to 295.4, while the starting states spanned from 20.7 to 41.6, and the three runs that ended without agent reported delivery were scored 19.0, 25.3, and 25.7. The launch-only run presented in Fig.~\ref{fig:launchonly} demonstrated above started at 20.7 and ended at 295.4.

We then tested the agent with a different language model (claude-opus-5) and a different tungsten tip. During the campaign on the second tungsten tip, the coarse piezo walker of the microscope jammed. Launched with a prompt that disclosed the jam, claude-opus-5 delivered the desired tip endpoint in 116~min of active operation (Fig.~\ref{fig:verification}(b)). The BraggZ score of the delivered topograph was 79.5 and the topograph was verified by human expert as a successful delivery. Its score was lower than other delivered topographs' scores because the lattice was resolved at low contrast due to a stronger FFT center background. This delivered endpoint also showed a $q_{\mathrm{AFM}}$ modulation in the Fourier transform (Fig.~\ref{fig:verification}(b)). This $q_{\mathrm{AFM}}$ modulation provides evidence that the delivered tungsten tip was spin-polarized. The jam caused the agent to change its strategy. It issued four smaller relocations: 300, 500, 200, and 200 steps, each followed by its verification. To compare, the run on the previously healthy walker had all relocations with more than a thousand steps. The two successive bias pulse interventions from this run produced opposite outcomes. The first turned the tip into a clustered multi-apex state, and the second sharpened it (Fig.~\ref{fig:interventions}). We therefore achieved launch-only end-to-end delivery with two frontier models, using STM tips made from different tungsten wires, and even with a degraded piezo walker. We then launched another tip conditioning task with a less intelligent LLM, gpt-5.4-mini. Under the same jam condition and providing the byte-identical launch prompt, the agent did not deliver the task. Its walks began at 200 and 500 steps and then ran repeatedly at 1000 to 3000, and its experiment log showed no other adjustment to the jam.

\begin{figure*}[!t]
\centering
\includegraphics[width=0.92\textwidth]{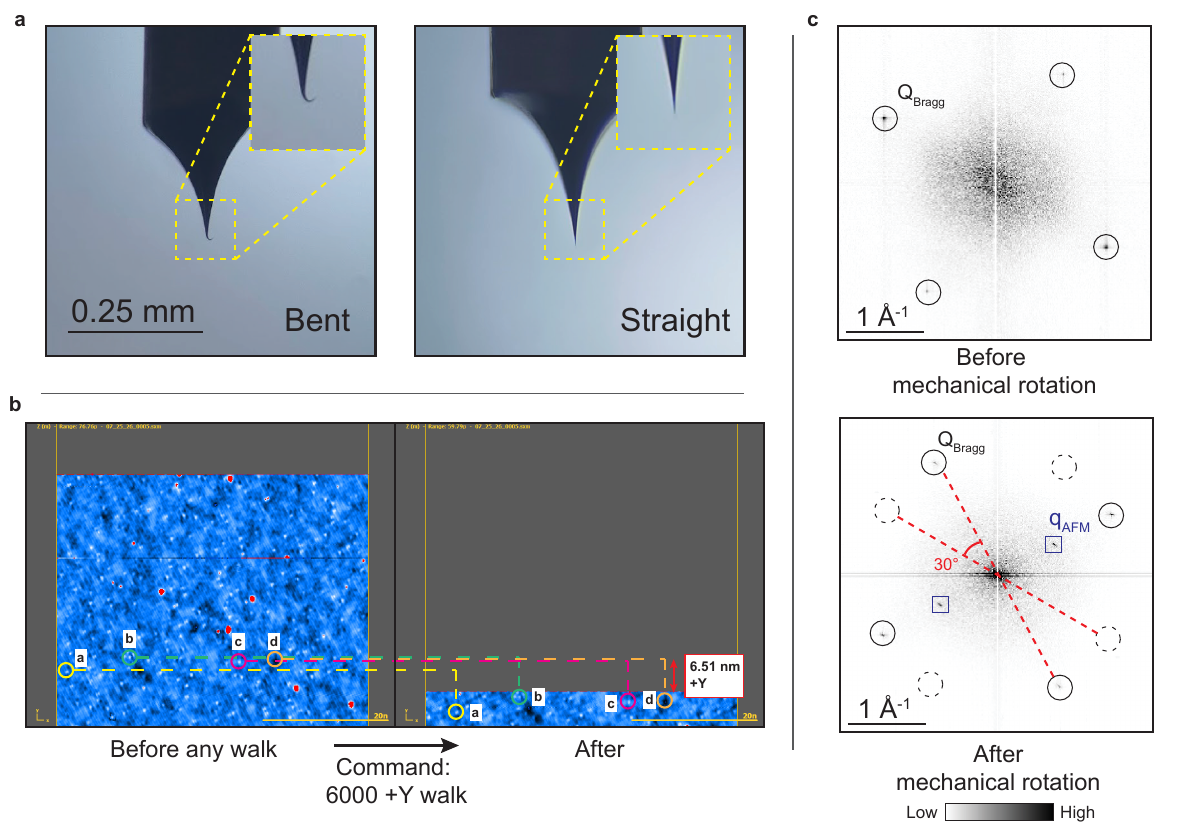}
\caption{\textbf{Unexpected conditions in STM tip conditioning.}
(a), Optical microscopic images of a bent (left) and a straight (right) tungsten tip.
(b), Processed live-scan images before (left) and after (right) commands for a total of 6,000 coarse-walker steps along $+Y$. Colored circles mark corresponding defect features. Dashed lines connect their positions. The red arrow marks the displacement of 6.51~nm along $Y$.
(c), Amplitude FFTs of topographs from the pre-rotation (top) and post-rotation (bottom) periods. Both topographs were acquired over a $50\times50$~nm field of view at a scan angle of $0^{\circ}$. Solid circles mark the first-order Bragg peaks. Dashed circles in the bottom FFT indicate the pre-rotation orientation. The fourfold Bragg pattern determines the lattice orientation only modulo $90^{\circ}$. The $120^{\circ}$ mechanical rotation therefore appears as the marked $30^{\circ}$ step. STM setup conditions: (c), $V_{sample}$ = 100 mV, $I_{set}$ = 200 pA.}
\label{fig:wild}
\end{figure*}

\subsection{Unexpected conditions and agent response}
We also examined the agent's behavior when it encountered unexpected conditions it had never met. In a real STM experiment, a bent tip usually results from accidental operations when transferring the tip from the ambient environment to the ultra high vacuum (UHV) chamber. The human operator intentionally bent a tungsten tip (Fig.~\ref{fig:wild}(a)), loaded it into the microscope, and did not tell the agent. The agent initially reported that the lattice was resolved, but the operator rejected that assessment. The agent continued to work and later reached a stable, conducting tip endpoint on a clean area. But the tip was not able to resolve the lattice. The agent stopped after trying all its available conditioning procedures. During this run the agent reported the tunneling current constantly reaching 3~nA, which is a sign of the tip hitting the sample directly. However, the topographs over this region showed only sparse junk. The current channel of the topograph showed that the current did not actually reach 3~nA when the tip scanned over the sparse junk. Instead, the 3~nA spikes appeared while scanning over a completely smooth area. The agent noticed this and suggested a stubborn second tunneling apex as the explanation, but it never realized the tip was bent. The human operator then took over the tip conditioning but also failed to obtain full lattice resolution within a reasonable intervention period (Fig.~\ref{fig:verification}(c)). This case shows that the agent can recognize persistent problems and stop an unsuccessful procedure, but it does not prove that the bent tip could be uniquely identified from the observations available to the agent.

A fresh, straight tungsten tip later replaced the bent tip, and a new agent run started. This run finished in 222~min and failed. Initially, the agent conditioned the tip and judged that it had improved, but the area was still not clean. After many piezo walks, no clean region appeared, and every reachable area remained full of impurities. The agent stopped and suggested replacing or mechanically reforming the tip, which required capabilities outside its workspace. The human operator then took over and rotated the sample holder by 120 degrees (Fig.~\ref{fig:wild}(c)), exposing a new region for the agent to explore. This rotation is a mechanical adjustment with no software control. It is not listed in the workspace file, nor was it part of the agent's apprenticeship or memory. The human operator then let the agent continue. The agent reported success after 29 minutes, using no bias pulses and only one mild shake. These observations suggest that the tip might have already been close to the required imaging condition before the rotation, and the main obstacle was simply finding a clean surface.

The piezo walker jam mentioned in the previous section is the other unexpected problem. When it first occurred, it was not disclosed to the agent. The agent noticed the abnormal behavior and guessed that the walker had carried the tip off the crystal onto the mounting epoxy, so it increased its travel commands accordingly. It later realized that knowing how many steps it had commanded did not tell it how far the walker had actually moved. The operator then told the agent about the jam, increased the drive voltage from 170 to 180 V (a common fix for STM piezo jams), and resumed the task. Although the piezo voltage can be controlled through the software, this capability was disabled in the agent's workspace for safety reasons. The agent then issued three successive commands of 2,000 coarse-walker steps each along +Y. Later analysis of the impurities across the live-scan images showed a displacement of only about 6.5 nm along Y (Fig.~\ref{fig:wild}(b)), confirming the jam. Despite this, the agent eventually reported that the task was complete, and the final topograph passed human expert verification.

\begin{table*}[!t]
\centering
\footnotesize
\setlength{\tabcolsep}{4pt}
\renewcommand{\arraystretch}{1.25}
\caption{\textbf{Unexpected conditions breakdown.} Each row is one reported case. Run identifiers, durations and message counts are listed in Table~S1 of the Supplemental Material~\cite{SM}.}
\label{tab:contingency-cases}
\begin{tabular}{p{2.5cm} p{4.9cm} p{3.7cm} p{3.8cm}}
\hline
Case & Agent Observation and Report & External Intervention & Summary \\
\hline
Bent tip & recurrent imaging abnormalities, current spikes, claimed lattice resolution, second tunneling apex proposed, stopped after conditioning options exhausted & human takeover but did not resolve the lattice & difficulty recognized, bent tip geometry not identified \\
\hline
Sample rotation & having trouble finding a suitable region, proposed tip replacement & manual 120 degree sample-holder rotation & delivery after physical intervention unavailable through the interface, tip may already have been near condition \\
\hline
Undisclosed walker jam & landings attributed to epoxy-like material or an off-crystal walk, later stated the record held commands and not physical position & jam disclosed, drive raised 170 to 180 V & abnormal behavior recognized, jam not diagnosed, recovery after disclosure and drive adjustment \\
\hline
Disclosed walker jam & Modified relocation strategy & jam disclosed at launch, no further external interventions & adaptation with the malfunction known from launch \\
\hline
Disclosed walker jam, weak-model run & no successful delivery, larger travel commands up to 3000 steps & jam disclosed at launch, one resumption message & disclosure alone did not lead to adaptation \\
\hline
\end{tabular}
\end{table*}

\section{Discussion}

Our experiments demonstrate the LLM agents can complete STM tip conditioning task within the \texttt{quailbot} harness. In addition, in Fig.~\ref{fig:verification}(b) the agent once delivered a spin-polarized tip endpoint, even though spin polarization was not an explicit objective of this task. This could suggest that an LLM agent can also systematically end-to-end prepare a tungsten tip into a spin-polarized state. In STM experiments, a scientifically reasonable action does not guarantee intended outcome. The tip apex state cannot be measured directly, and the effect of an intervention is often uncertain. The agent must infer the current experimental state from incomplete evidence, check the effects of its interventions, and revise its strategy accordingly. In the launch-only run shown in Fig.~\ref{fig:launchonly}, experimental readbacks twice contradicted the agent's expectations, prompting it to change its next action without human intervention. The two successive bias pulse interventions in Fig.~\ref{fig:interventions} further demonstrate that tip conditioning can be unpredictable, producing contrasting imaging outcomes, one degrading and the other improving image quality. The outcome of such an intervention is not known in advance, so a scripted sequence automation cannot suffice.

Beyond demonstrating successful end-to-end LLM operation on an STM, our experiments provide additional insight on why LLM agents can fail in real physical experiments. These failures cannot always be explained by insufficient scientific knowledge or inadequate experimental planning. Recent autonomous-experiment studies have reached related conclusions from different directions. In trapped-ion experiments~\cite{wang2026artiq}, human guidance was still needed when the agent failed to recognize that an unsuccessful approach needed to be reconsidered. An another autonomous NV-center experiment~\cite{isogawa2026agenticquantumsensing} similarly showed that an agent could miss the effect of a residual resonance-calibration offset on a subsequent Ramsey experiment, and called for a dedicated checkpoint benchmark to test whether agents recognize this calibration problem.

The contingency cases expose the limits of the agent's demonstrated capability -- recognizing an anomaly did not always lead to identifying its cause or recovering from it (Table~\ref{tab:contingency-cases}). In the successful run with the walker jam disclosed at launch, the agent adjusted its motion strategy using that information. On the other hand, in the bent-tip and undisclosed walker-jam cases shown in Fig.~\ref{fig:wild}, the agents recognized abnormal behavior but did not identify its cause from the available observations. The agents relied on explanations learned during apprenticeship, such as multiple tunneling apices, surface contamination, or movement onto epoxy. In the walker case, the agent did not identify the jam, but it eventually realized that knowing how many steps it had commanded did not tell it how far the walker had actually moved. It therefore could not be sure where the tip was. Taken together with the residual-calibration example reported in autonomous NV experiments, these cases illustrate several reasons why an agent may fail to identify the underlying condition: the evidence may be indirect, the necessary information may be unavailable through the instrument interface, or the agent may not consider the correct explanation. In our experiments, human input was needed mainly when these unexpected conditions arose.

Autonomy also depends on what the instrument allows the agent to observe and control. An agent cannot distinguish physical states that produce the same readbacks, or perform an action that its interface does not support. For example, in the sample-holder rotation case, the tip may already have been close to the required imaging condition, but the agent could not find a suitable clean region. Exposing a fresh region required mechanical rotation of the sample holder, which could not be controlled through software. Human operators can compensate for such gaps using information outside the digital control channel, such as visual inspection and direct physical intervention. Agent-ready scientific instruments should therefore provide access to essential control functions and diagnostic information, keep a clear record linking actions to their measured outcomes, and support automated verification and recovery. The Model Hardware Standard~\cite{anthropic2026mhs} provides a common way for agents to read instrument data and issue commands. However, a software interface can only expose measurements and controls that the instrument actually provides. Additional sensors or actuators may be needed to give the agent information or capabilities currently available only to a human operator. Nevertheless, interface standardization does not by itself reveal the atomic structure of the tip apex or predict how a conditioning pulse will change it.

Real-world scientific agents are beginning to reveal a family of failure modes that conventional LLM benchmarks largely collapse into 'reasoning failures'. Our experiments expose a class of problems that becomes unavoidable once a LLM itself enters the experimental feedback loop. Evaluations should therefore test more than whether an agent can complete a workflow under expected conditions. They should examine how agents respond when experimental states cannot be measured directly, interventions have uncertain or history-dependent effects, unexpected conditions arise, or a necessary action is unavailable through the instrument interface. They should also assess whether an agent recognizes when it needs additional information or human intervention.

\section{Methods}
\subsection{Instrument}

The experiments were performed on a customized Unisoku USM1300 low-temperature STM at a base temperature of 4.5~K. The STM tips were home-made chemically etched tungsten tips, annealed in an ultrahigh-vacuum chamber to a bright orange color before the experiments. 

\subsection{Bragg-peak evaluation}

We evaluated the raw topographs after each experiment to compare the initial and final STM states. We first removed the slowly varying background from each scan line. We then removed the image mean, applied a two-dimensional Hann window, and calculated the amplitude of the Fourier transform. The windowed Fourier data were used only for scoring. The unwindowed Fourier images shown in the figures were generated separately.

We searched the Fourier amplitude for the first-order Bragg rings expected from the Te and Fe lattice spacings. For each ring, we identified four symmetry-related Bragg peaks and compared their combined intensity with the surrounding intensity at the same wave vector. We selected the ring with the stronger four-peak signal. BraggZ is the mean excess peak intensity normalized by the median absolute deviation of the surrounding background,
\[
\mathrm{BraggZ}=\frac{1}{4}\sum_{i=1}^{4}\frac{S_i-B}{D},
\]
where $S_i$ is the local amplitude around peak $i$, $B$ is the median background amplitude, and $D$ is the unscaled median absolute deviation of that background. BraggZ measures the contrast of the expected lattice peaks relative to the Fourier background. It is not a statistical significance or a delivery threshold. We used the score together with inspection of the real-space topograph and Fourier image when evaluating the experimental endpoint.

\subsection{Harness implementation}

\texttt{Quailbot} is implemented as a Pi\cite{piagent} extension. The workspace JSON declares instrument parameters, named actions, permitted operations, linked observables, and the configured instrument driver. The harness registers four driver-agnostic instrument-control CLI tools: \texttt{cli\_get}, \texttt{cli\_set}, \texttt{cli\_ramp} and \texttt{cli\_action}. \texttt{cli\_get} is read-only, and its result lists the capability's declared linked observables. \texttt{cli\_set}, \texttt{cli\_ramp} and \texttt{cli\_action} are state-changing calls. The tools are instrument-agnostic. We released \texttt{quail-cli-core}\cite{quailclicore}, a driver kit for wrapping an API-exposing controller in this schema. In our experiments, we use \texttt{nspmctl}\cite{nspmctlcli} to wrap the Nanonis SPM Controller\cite{nanonisspm} driver of our low-temperature STM. The harness resolves every call against the workspace before invoking the driver. State-changing tools are denied by default: \texttt{cli\_set}, \texttt{cli\_ramp} and \texttt{cli\_action} run only when the human operator enables them with the \texttt{QUAILBOT\_ALLOW\_MUTATING\_TOOLS} environment flag. A denied call is rejected before it reaches the instrument, and the denial is recorded in the experiment log. For a state-changing call, it attempts the linked observables declared for that capability after execution. The tool result contains the primary driver result and the linked-observation records. The context hook injects this result into the agent context before the next model turn. The experiment log stores the tool name, executed arguments, primary result, linked-observation outcomes and linked image files. Failed linked-observable acquisitions remain in the record as failures. In this paper, the experiment log refers to these harness-generated records, the per-run \texttt{events.jsonl} file and its linked observation blobs. The human operator's lab notebook is a separate record. 

\subsection{Memory and skills implementation}
The memory and skills modules serve different purposes. Memory holds empirical knowledge, for example, a representative entry records a motor-walk sequence that is safe on this FeTe sample and dangerous on another sample. Skills, on the other hand, hold general procedures, such as the tip conditioning techniques. The released skill set comprises sample exploration, tip conditioning, tip pulsing, and hidden-state inference. Skills are designed to be progressively disclosed, only names and one-line descriptions stay permanently in the agent's context, and the agent loads the full body on invocation. Both memory and skills are evolving modules. The agent was taught in its system prompt to reflect on its actions and their outcomes, and consolidate what it learned by updating its memory and skills in flight. At the end of the experiments the memory held 40 entries containing general rules and local calibration values that organize STM control around public states, hidden states, and hidden dynamics.

\begin{acknowledgments}
Siyu Cheng thanks Jiaqi Cai for helpful discussion, GaugeForge PTE LTD for part of technical support, and Genda Gu at BNL for the FeTe sample. This work was supported by the Gordon and Betty Moore Foundation, grant DOI 10.37807/GBMF12956. We acknowledge the use of Codex (OpenAI) for assistance in developing the \texttt{quailbot} software.
\end{acknowledgments}

\section*{Author contributions}
S.C. developed \texttt{quailbot}, performed the STM experiments and collected the agent trajectories. M.X. and C.C. participated in training the \texttt{quailbot} agent. I.Z. and S.C. wrote the paper with the input from all the authors. I.Z. supervised the project.

\section*{Data availability}
The raw STM data (SXM files) and the agent records from July 15 to July 27, 2026 will be deposited in Zenodo upon publication. The agent records comprise the agent transcripts including the task prompts, the harness-generated experiment logs and the linked observation images. The \texttt{quailbot} source code is available on GitHub~\cite{quailbotpi}.

\bibliography{bib/references}

\end{document}